
\magnification=1200  
%
\def\newline{\hfill\penalty -10000}  
\def\title #1{\centerline{\rm #1}}
\def\author #1; #2;{\line{} \centerline{#1}\smallskip\centerline{#2}}
\def\abstract #1{\line{} \centerline{ABSTRACT} \line{} #1}
\def\heading #1{\line{}\smallskip \goodbreak \centerline{#1} \line{}}
\newcount\refno \refno=1
\def\refjour #1#2#3#4#5{\noindent    \hangindent=1pc \hangafter=1
 \the\refno. #1, #2 ${\bf #3}$, #4 (#5). \global\advance\refno by 1\par}
\def\refbookp #1#2#3#4#5{\noindent \hangindent=1pc \hangafter=1
 \the\refno.  #1, #2 (#3, #4), p.~#5.    \global\advance\refno by 1\par}
\def\refbook #1#2#3#4{\noindent      \hangindent=1pc \hangafter=1
 \the\refno.  #1, #2 (#3, #4).           \global\advance\refno by 1\par}
\newcount\equatno \equatno=1
\def\adveqn{(\the\equatno) \global\advance\equatno by 1}

\def\up#1{\leavevmode \raise 0.2ex\hbox{#1}}

\newcount\figno
\figno=0
\def\figure{\global\advance\figno by 1 Figure~\the\figno.~}
%

%
\vsize=8.75truein
\hsize=5.75truein
\hoffset=0.5truein
%
\baselineskip=0.166666truein
%
\parindent=25pt
%
\parskip=0pt
%
\nopagenumbers

\input tables

\def\sles{\lower2pt\hbox{$\buildrel {\scriptstyle <}
   \over {\scriptstyle\sim}$}}
\def\sgreat{\lower2pt\hbox{$\buildrel {\scriptstyle >}
   \over {\scriptstyle\sim}$}}

\def\ap{\approx}
\def\T{\tau}

\def\g{\gamma}
\def\a{\alpha}

\def\st{\sigma_T}

\def\R07{R_{i,7}}
\def\ef{\eta_4}
\def\E051{E_{i,51}}
\def\dR{\Delta R}

\def\g{\gamma}
\def\e{\eta}

\line{}
\title{\bf FIREBALLS}
\author
Tsvi Piran\parindent=0pt  ;
Racah Institute for Physics, The Hebrew University, Jerusalem 91904, Israel ;

\abstract{
The sudden release of copious $\g$-ray photons into a compact region
creates an opaque photon--lepton fireball due to the prolific
production of electron--positron pairs.
The photons that we observe in the bursts
emerge only at the end of the fireball phase after it expanded
sufficiently to become optically thin or after it converted its
energy to the kinetic energy of relativistic baryons which convert it,
in turn, to electromagnetic pulse via the interaction with
interstellar matter. It is essential, therefore, to analyze the
evolution of a fireball in order to comprehend the observed features of
$\gamma$-ray bursts. We discuss various aspects of fireball
hydrodynamics and the resulting emitted spectra. }

\heading{1. Introduction - The Inevitability of Fireballs}

The recent observation of the BATSE experiment on the COMPTON-GRO
observatory  have demonstrated, quite convincingly that $\gamma$-ray
bursts (grbs) originate from cosmological sources$^{1,2}$.
Preliminary evidence for the predicted$^{3,4}$ correlations between
the duration the hardness, the strength and the hardness of the
bursts$^{5,6}$ supports this conclusion. The correlation suggests,
in agreement with an analysis cosmological $C/C_{min}$ distribution$^{4}$
that the weakest bursts originate from distances of $z \approx 1$,
corresponding to a release of  $ E \approx 10^{51}$ergs (if the
emission is isotropic).

The rapid rise time observed in some of the bursts implies that the
sources are compact and that in some cases the size of the source
$R_i$ is as small as 100km.  The copious release of energy within such
a small volume results in an initially optically thick system of
photons, electrons and positrons which we call a ``fireball".  The term
``fireball" refers here to an opaque radiation - plasma whose initial
energy is larger than its rest mass.  The initial optical depth in
cosmological grbs for $\g \g \rightarrow e^+ e^- $ is$^{8}$:
$$
\tau_{\gamma\gamma} = {f_g E \st / R^2 m_e c^2} \approx 10^{19} f_g
\E051 \R07^{-2} ,
\eqno(e1)
$$
where $\E051$ is the initial energy of the burst in units of
$10^{51}$ergs, $\R07$ is the radius into which the energy is injected
in units of $10^7$cm and $f_\g$ is the fraction of primary photons
with energy larger than $2m_e c^2$.  Since $\tau_{\g\g} \gg 1$ the
system reaches rapidly thermal equilibrium (regardless of the initial
energy injection mechanism) with a temperature: $T = 6.4 \E051^{1/4}
\R07^{-3/4}$MeV. At this temperature there is a copious number of
$e^+-e^-$ pairs which in turn contribute to the opacity via Compton
scattering.

The huge initial optical depth prevent us from observing directly the
radiation released by the source regardless of the specific nature or
the source.  The observed radiation emerges only after the fireball
has expanded significantly and became optically thin.  We should
divide, therefore, the discussion of cosmological grbs to two parts:
the nature of the energy source (in another paper in this volume$^{8}$ we
discuss the binary neutron star merger model) and the evolution of a
fireball (which we address here).  The the fireball
phase determines the observational features of grbs.
This can be compared to the situation in stars in which energy is
generated in the core but it leaks out to through an optically thick
envelope and the observed spectrum is determined
by the conditions at the photosphere.  Similarly the
observed grb spectra is determined by the way that fireballs
evolve and release their energy.

Before turning to a discussion of the fireball evolution we discuss
two recent proposals to avoid fireballs in grbs.  The first idea
resembles the fireball to some extend as it is based on a
relativistic motion of the source.  The  observed photons are blue
shifted to $\gamma$-rays because of the relativistic motion of the
source.  The local temperature is much lower and  the fraction $f_g$
of high energy photons at the source is sufficiently small that
$\tau_{\g\g}$ would be less than one and the photons would escape
freely (this is the case, for example, in the pure radiation fireball
that we discuss latter).  The opacity of a source with a spectral
index $\a$ moving toward the observer with a relativistic factor $\g$
is$^{9}$:
$$
\T_{\g \g} \ap 6.5\times 10^6 (2\g)^{- (2+\a )} R_6
A_{12}^{-2}F_{-7} D_{100}^2 \ ,
\eqno(2)
$$
where $F = F_{-7} \times 10^{-7}~ {\rm erg sec}^{-1}{\rm cm}^{-2}$ is
the observed flux, $A = 10^{12} \times A_{12}~ {\rm cm}^2$, is the
area of the emitting region and $D = 100 \times D_{100}$kpc is the
distance to the source.  Even at the galactic halo one requires $\g >
30$ for $ \a = 2$ or $\g > 100 $ for $\a = 1$.  This solution raises,
therefore, several other problems which are as serious as the problem
that it solves: What is the accelerating mechanism that accelerates
the grb sources to such a high relativistic velocities?  What is the
source of the huge kinetic energy required for the bulk motion of such
sources?

Alternatively, a fireball will not appear if the energy is releases
non-electromagnetically and it is converted to photons at significantly
larger distance in which eq. 1 yields $\T_{\g\g} \ll 1$.  This can
happen, for example, if the source emits weakly interacting particles
which are somehow converted in route to photons. Such a model based on
emission of axions by supernova has been recently suggested$^{10}$.
It does not explain, however, how to reconciles the supernova rate and
the grb rate (which differ by a factor of a thousand) ? and how can GeV
photons$^{11}$ emerge from such sources?

\heading{2. Fireball Evolution - an Overview}

Consider, first, a pure radiation fireball.  Initially, when the local
temperature $T$ is large, the opacity is large due to $e^+e^-$
pairs$^{12}$ and the radiation cannot escape freely.  The fireball
expands and cools and this opacity, $\tau_p$, decreases exponentially
with decreasing temperature.  At $T_p\approx 20$ KeV, $\tau_p \approx
1$, the fireball becomes transparent, the photons escape freely and
the fireball phase ends. While the local temperature is $T_p$ the photons
are blue shifted roughly to the original
temperature due to the relativistic motion of the fireball at that stage.
Preliminary calculations$^{12,13,14}$ show that
unlike the spectra  observed in grbs (see however $^{15}$) the
spectra emitted from a pure radiation fireball is a blended thermal spectrum.

In addition to radiation and $e^+e^-$ pairs, astrophysical fireballs
may also include some baryonic matter which may be injected with the
original radiation or may be present in an atmosphere surrounding the
initial explosion$^{13,16,17}$.  This affect the fireball in two ways: The
electrons associated with this matter increase the opacity, delaying
the escape of radiation.  More importantly, the baryons are
accelerated with the rest of the fireball and convert part of the
radiation energy into bulk kinetic energy.

As a loaded fireball with a baryonic mass, $M$, evolves two important
transitions take place.  One transition corresponds to the change from
optically thick to optically thin conditions.  The opacity itself has
a contribution from electron-positron pairs as well as electrons
associated with the baryons. Initially, when the local temperature $T$
is large, the opacity is dominated by $\tau_p$.  However, the matter
opacity, $\tau_b$, decreases only as $R^{-2}$, where $R$ is the radius
of the fireball.  Generally, at the point where $\tau _p =1$,
$\tau_{b}$ is still $ > 1$ and the final transition to $\tau=1$ is
delayed and occurs at a cooler temperature.  The photons escape freely
at this stages.  The electrons and the baryons are however still
coupled to the photons until the mean free path for a Compton
scattering of an electron on a photons drop to unity. This happens at
a slightly larger radius.

The second transition corresponds to the switch from radiation
dominated to matter dominated conditions, i.e from $\e >1$ to $\e <1$,
where $\e \equiv E/Mc^2$, the ratio of the radiation energy $E$ to the
rest energy $M$.  In the early radiation dominated stages when $\e>1$,
the fluid accelerates in the process of expansion, reaching
relativistic velocities and large Lorentz factors.  The kinetic energy
too increases proportionately.  However, later when $\e < 1$, the
fireball becomes matter dominated and the kinetic energy is comparable
to the total initial energy.  The fluid therefore coasts with a
constant radial speed.  The overall outcome of the evolution of a
fireball then depends critically on the value of $\e$ when $\tau$
reaches unity (or equivalently on whether $R_\e > R_\tau$ or vice
versa).  If $\e>1$ when $\tau=1$ most of the energy comes out as high
energy radiation, whereas if $ \e < 1 $ at this stage most of the
energy has already been converted into kinetic energy of the baryons
and we have to examine the fate of those extreme relativistic baryons.

The initial ratio of radiation energy to mass, $\e_i$, determines in
what order the above transitions take place.  Shemi and
Piran$^{13}$  identified four regimes:

\noindent
{(i)} $\e_i > \e_{pair} = (3 \sigma_T^2 E_i \sigma T_p^4 /4 \pi m_p^2 c^4 R_i
)^{1/2} \approx 10^{10} \E051^{1/2} \R07^{-1/2}$
(corresponding to $M< M_{pair} =5 \times 10^{-13} m_\odot) \E051^{1/2}
\R07^{1/2}$): In
this regime the effect of the baryons is negligible and the evolution
is of a pure photon-lepton fireball.  When the temperature reaches
$T_p$, the pair opacity $\tau_p$ drops to 1 and $\tau _b \ll 1$.  At
this point the fireball is radiation dominated ($\e >1$) and so most
of the energy escapes as radiation.

\noindent
{(ii)} $\e_{pair} > \e_i > \e_{b} =
(3 \sigma_T E_i / 8 \pi m_p c^2 R_i^2)^{1/3} \approx
10^5 \E051^{1/3}  \R07^{-2/3}$ (corresponding to $M_{pair}< M < m_b=
5 \times 10^{-8} m_\odot \E051^{2/3}  \R07^{2/3})$: Here, in the late
stages, the opacity is dominated by free electrons associated with the
baryons.  The comoving temperature therefore decreases far below $T_p$
before $\tau$ reaches unity.  However, the fireball continues to be
radiation dominated as in the previous case, and most of the energy
still escapes as radiation.

\noindent
{(iii)} $\e_{b} > \e_i > 1$ (corresponding to $M_b < M <
5 \times 10^{-4} m_\odot \E051$):
The fireball becomes matter dominated before it becomes
optically thin.  Therefore, most of the initial energy is converted
into bulk kinetic energy of the baryons, with a final Lorentz factor
$\g_f \approx  \e_i$.

\noindent
{(iv)} $\e_i < 1$: This is the Newtonian regime.  The
rest energy exceeds the radiation energy and the expansion never
becomes relativistic. This is the situation, for example in supernova
explosions in which the energy is deposited into a massive envelope.

\heading{3. Extreme Relativistic Scaling Laws}

After an initial acceleration phase in which the fireball reaches
relativistic velocities and $\g \sgreat {\rm few}$ each shell of an
extreme relativistic fireball satisfies to order $o(\g ^{-2})$ the
following conservation laws$^{14}$:
$$
r^2n\g = {\rm const.}, \qquad r^2e^{3/4}\g = {\rm const.},
\qquad r^2  (n+4e/3)^2 = {\rm const.},
\eqno (3)
$$
where $n$ and $e$ are the local baryon and energy densities.  These
scalings were derived for a homogeneous radiation dominated
fireball$^{13,12}$ by noting the analogy with an expanding universe.
The same relations are valid, however, for each individual radial
shell in the fireball even in the more general inhomogeneous case.
These scaling laws also apply to Paczy\'nski's$^{18}$ solution for a
steady state relativistic wind. They are
valid even for fractions of individual shells provided that some
general conditions on the angular motion are satisfied.

Eqs. 3 yields a scaling solution which is valid everywhere provided
that $\g \sgreat  few $.  Let $t_0$ be the time and $r_0$ be the radius at
which a fluid shell in the fireball first becomes ultra-relativistic,
with $\g \ \sgreat\ {\rm few}$.  Label various properties of the shell
at this time by a subscript $0$, e.g. $\g _0$, $n_0$, $e_0$, and
$\eta _0 = e_0/n_0$.  Defining the auxiliary
quantity $D$, where
$$
{1 \over D} \equiv
{\g_0 \over \g } + {3\g_0 \over 4\eta_0 \g } - {3\over 4\eta_0},
\eqno(4)
$$
we find that
$$
r = r_0 { D^{3/2} (\g_0 / \g)^{1/2}},
\qquad n = n_0  D^{-3},
\qquad e = e_0  D^{-4},
\qquad \e = \e_0  D^{-1}.
\eqno (5)
$$
These are parametric relations which give $r,~n,~e$, and $\e$ of each
fluid shell at any time in terms of the $\g$ of the shell at that
time.  The relation for $r$ in terms of $\g$ is a cubic equation.
This can in principle be inverted to yield $\g(r)$, and thereby
$n,~e$, and $\e$ may also be expressed in terms of $r$.

The parametric solution 5 describes both the radiation-dominated
and matter-dominated phases of the fireball within the frozen pulse
approximation.  For $\g\ll \e _0\g _0$, the first term in eq. 4
dominates and we find $D\propto r$, $\g\propto r$, which yields the
radiation-dominated scalings of eqs. e7.  This regime extends out to a
radius $r\sim \e_0 r_0$.  At larger radii, the first and last terms in
eq. 4 become comparable and $\g$ tends to its asymptotic value of $\g_f
= (4\e_0/3+1)\g_0$.  This is the matter dominated regime.  (The
transition occurs when $4e/3=n$, which happens when $\g=\g_f/2$.)  In
this regime, $D\propto r^{2/3}$, leading to the matter dominated
scalings laws (eqs. 10).

It is unlikely that a realistic fireball will be spherically
symmetric. In fact strong deviation from spherical symmetry are
expected in the most promising neutron star merger model, in which the
radiation is expected to emerge through funnels along the rotation
axis$^{8}$.  The initial motion of the fireball might be
fairly complex but once $\g \gg 1$ and provided that some
some simple conditions are satisfied then
the motion of each fluid element
decouples from the motion of its neighbors and it  can be described by
the same asymptotic solution, as if it is a part of a spherical shell.
We define the spread angle $\alpha$ as $ u^r \equiv u \cos\alpha $ and the
angular range over which different quantities vary as $\Delta
\theta$. Eqs. 3 hold locally if:
$$
\alpha < \Delta \theta \ \ \ \ {\bf \rm and }\ \ \ \
( \alpha < 1/\gamma \ \ \ \ {\rm or}\ \ \ \   1/\g < \alpha \ll 1 ).
\eqno(6)
$$

\heading{4. Physical Conditions in the Fireball}

\heading{\it 4.i Radiation-Dominated Phase}

The fireball is initially radiation-dominated. During this phase
($e\gg n$) and:
$$
\g\propto r,\qquad n\propto r^{-3},\qquad e\propto r^{-4},
\qquad T_{obs}\sim {\rm constant},
\eqno (7)
$$
where $T_{obs}\propto \g e^{1/4}$ is the temperature of the radiation
as seen by an observer at infinity.  (Strictly, the radiation
temperature depends on $e_r$, the energy density of the photon field
alone; for $T \ll m_ec^2$, $e_r=e$, but for $T> m_e c^2$, $e$ contains
an additional contribution from the electron position pairs$^{13}$ we
neglect this complication for simplicity). The scalings of $n$ and $e$
given in eqs. 7 correspond to those of a fluid expanding uniformly in
the comoving frame.  Although the fluid is approximately homogeneous
in its own frame, because of Lorentz contraction it appears as a
narrow shell in the observer frame, with a radial width given by:
$$
\Delta R \sim R/\g \sim {\rm constant}\sim R_i  \ \ \ .
\eqno(8)
$$
We interpret eq. 7 and the constancy of the radial width $\Delta r$
in the observer frame to mean that the fireball behaves like a pulse
of energy with a frozen radial profile, accelerating outward at almost
the speed of light.

If there are no baryons this phase last until the local temperature
drops to $T_p$ and the fireball becomes optically thin.
Preliminary calculations$^{12,13,14}$ show that
unlike the spectra  observed in grbs (see however $^{15}$) the
spectra emitted at this stage is a blended thermal spectrum.

If baryons are present the radiation dominated phase
lasts from the initial size, $R_i$, until $\g = \eta$ at $R_\eta$
$$
R_\eta = 2 R_i \eta  = 2 \times 10^{11} {\rm cm} \  \R07 \ef,
\eqno(9)
$$
where the initial thermal
energy is converted to the kinetic energy of the baryons:

\heading{\it 4.ii Matter-Dominated Phase}

In the alternate matter-dominated regime ($e\ll n$), we obtain from
eq. 3 the following different set of scalings,
$$
\g\rightarrow {\rm constant},\qquad n\propto r^{-2},
\qquad e\propto r^{-8/3},\qquad T_{obs}\propto r^{-2/3}.\eqno
(10)
$$
The modified scalings of $n$ and $e$ arise because the fireball now
moves with a constant radial width in the comoving frame.  (The
steeper fall-off of $e$ with $r$ is because of the work done by the
radiation through tangential expansion.)  Moreover, since $e\ll n$,
the radiation has no important dynamical effect on the motion and
produces no significant radial acceleration.  Therefore, $\g$ remains
constant on streamlines and the fluid coasts with a constant
asymptotic radial velocity.  The width of the fireball remains constant with:
$$
\dR = R_i .
\eqno(11)
$$

Eventually, as the particle density decreases this phase ends when
the electrons decouple from the photons at $R_c$:
$$
R_c = (a^{1/8}/k^{1/2} ) \sigma_T^{1/2} (4 \pi /3)^{-3/8} E_i^{3/8}
R_i^{3/8}  =  1.6 \times 10^{15}{\rm  cm} \  \E051^{3/8}  \R07^{3/8} .
\eqno(12)
$$

\heading{\it 4.iii Free Coasting}

At very late times in the matter-dominated phase the frozen pulse
approximation breaks down.  At $R \approx R_c$ the electrons
decouple from the photons and at $R>R_c$ the baryons, electrons and
photons coast freely. The spread in the Lorentz factor of the
baryons leads to a spreading of the fireball whose width becomes:
$$
\dR = R_i + {R / \gamma^2} = 10^7 ( \R07 + {R_{15} / \ef^2})
\eqno(13)
$$
The second term, that expresses the additional spreading, is comparable
to the original width at
$$
R_w \approx  \eta^2 R_i \approx 10^{15} {\rm cm} \
\ef^2 \R07
\eqno(14)
$$
However if $R_w < R_c$ the spreading does not begin until $R = R_c$.
$$
\dR \approx \cases{ {R /\eta^2} \approx 10^7{\rm cm} \ R_{15} \ef^{-2}
&~~~~ for $R>R_w$ and $R>R_c$ \cr\noalign{\smallskip}
R_i=
10^7{\rm cm} \ \R07 &~~~~ otherwise } .
\eqno(15)
$$

\heading{5. Interaction of the Fireball with the ISM}

M\'es\'zaros, and Rees$^{19,20}$ suggested that the
interaction between the ultrarelativistic baryons and the
interstellar matter (ISM) provides a way to convert back the kinetic
energy of the baryons to electromagnetic energy. The
situation is similar to the one in supernova remnants (SNRs) in which
the kinetic energy of the ejecta is converted to radio emission due to
interaction with the ISM. The mean free path of a relativistic baryon in
the ISM is $\sgreat 10^{26}$cm, hence the interaction between
the baryons and the ISM cannot be collisional. However, from the
existence of SNRs we can infer that a collisionless shock can form
(possibly via magnetic interaction).

The interaction  becomes significant at $R_\gamma$ where the fireball
sweeps an external mass of $M_0/\g_F=(E_i/\eta c^2)/\gamma_F$ and looses
half of its initial momentum:
$$
R_\gamma = \left [ M_0 \over (4 \pi / 3) n \gamma_F \right ]^{1/3}
= 1.3 \times 10^{15}{\rm cm} \ \E051^{1/3} \ef^{-2/3} n^{-1/3}
\eqno(16)
$$
Because of a numerical coincidence $R_w \approx  R_\gamma \approx  R_c$
for our canonical parameters.
$R_c$ is independent of $\eta$ while $R_w$  increases with $\eta$ and
$R_\g$ decreases with $\eta$. Therefore, $R_\g < R_c< R_w$
for $\eta > 10^4$ and  $R_w < R_c < R_\g$ for $\eta < 10^4$.

Just like in SNRs the interaction between the fireball and the ISM
produces a forward moving
shock, which propagates into the interstellar matter, and a reversed
shock propagating into the fireball (see Fig. 2).  Following
Katz$^{21}$ we denote as region 1 the interstellar
matter, $n_1=n$ and $e_1 \ll n_1 m c^2 $. Regions 2 and 3 describe the
shocked material.
Pressure equilibrium along the contact discontinuity between 2 and 3
requires $e= e_2 = e_3$.  Region 4 denotes the fireball where
$n_4 m c^2 \gg e_4$.

A critical parameter is f, the ratio of densities (in the local
fluid's frame) between region 4 (the fireball's material) and region 1
(the external matter):
$$
f = {n_4 \over n_1} =  { M \over [ (4 \pi) R^2 \dR \gamma ] n_1 } =
{ 1\over 3}  \left ( {R_\g \over R} \right)^2 \left ( {R_\g \over \dR }
\right ) \approx
\eqno(17)
$$
$$
\approx
\cases { 5 \times 10^{7}  \E051 \ef n^{-1} R_{15}^{-3}
&~~for $R>R_w$ and $R>R_c$ \cr\noalign{\smallskip}
5 \times 10^{7}  \E051 \ef^{-1} n^{-1} \R07^{-1} R_{15}^{-2}
&~~otherwise \cr}
$$
where $R_{15}$ is the radius in units of $10^{15}$cm.

The shock conditions between 1 and 2 yield$^{22,23}$:
$$
\gamma_{1,2} = 0.5  \sqrt{{e /  n_1 m_p c^2 }}
\ \  ;  \ \ n_2 = 4 \gamma_{1,2} n_1,
\ \  ;  \ \  e \equiv e_2 = \gamma_{1,2} n_2 m c^2
\eqno(18)
$$
where $\gamma_{1,2}$ is the Lorentz factor of the motion of the shocked
fluid relative to the rest frame of an external observer.

The  Lorentz factor of the shock front itself is $\sqrt{2} \gamma_{1,2}$
Similar relations hold for the reverse shock (with 3,4 replacing 1,2).
The definition of $f$ yields:
$ \gamma_{3,4} = f^{-1/2} \gamma_{1,2}$ and $ n_3  = f^{1/2} n_2 $.
Using this we can express $\g_{1,2}$ and $\g_{3,4}$  in terms of $\g_F$:
$$
\gamma_{1,2}  =  f^{1/4} \gamma_F^{1/2} /\sqrt 2
\ \ \ ; \ \ \ \gamma_{3,4} = f^{-1/4} \gamma_F^{1/2} /\sqrt 2 .
\eqno(19)
$$
This holds if
$f < \gamma_F^2 \approx \eta^2$. Otherwise the reverse shock is not
relativistic and:
$$
\gamma_{1,2}  \approx \gamma_F
\ \ \ ; \ \ \ \gamma_{3,4} \approx 1.
\eqno(20)
$$

Since $f$ decreases with $R$ the reverse shock is initially non
relativistic.  Using eq. 17 we find that $f(R_\g) < \eta^2$ only if
$R_w > R_\g$. In this case $f(R_\g) \approx (\eta^2/3)(R_\g/R_w)$ and
a mildly relativistic reverse shock develops, with $\g_{3,4} (R_\g)
\approx (R_w/R_\g)^{1/4}$.  If $R_\g >R_w$ and $R_\g>R_c$,
$f=\eta^2/3$ at $R_\g$. $f$ decreased with $R$ and one might expect
that a relativistic reverse shock will develop latter.  The reverse
shock reaches, however, the inner boundary of the fireball shell when
the fireball reaches $R_\g$ and a rarefraction wave begins to move
forwards from the back of the fireball before a relativistic reverse
shock develops.

\heading{6. Energy Generation via Synchrotron  Cooling}

The kinetic energy of the fireball is converted to thermal energy at
the shocks. This happens in an optically thin region and the resulting
photons can escape and produce the observed grbs. The most likely
mechanism for the conversion of the thermal energy to kinetic energy
is via synchrotron cooling of the ultra-relativistic electrons.  This
mechanisms requires a strong coupling between the electrons, which
radiate the energy, and the protons, that carry the kinetic
energy. It also requires a strong magnetic field.
M\'es\'zaros and Rees$^{24}$ discuss various emission mechanisms from
the shocks. We discuss here an example of the simplest model.

We assume  equipartition  between the magnetic and the thermal energies.
Using eq. 18 we find:
$$
B = .5 {\rm Gauss} \gamma_{1,2} n_1^{1/2}
\qquad {\rm and} \qquad
\epsilon_L = 10^{-8} eV   \gamma_{1,2} n^{1/2} ,
\eqno(21)
$$
where $\epsilon_L$ is the  corresponding  Larmour energy.  Assuming
equipartition between the kinetic energy of the shocked electrons and
the shocked protons the typical Lorentz factor of the electrons
is larger by $(m_p/m_e)$ then the Lorentz factor of the protons.
The typical  energy of an emitted photon
is $\epsilon_{synch} = (m_p/m_e)^2 \gamma_{1,2}^2
\epsilon_l$. This is blue shifted by another factor of $\g_{1,2}$ for
an observer an rest.  Thus:
$$
\epsilon_o \approx 4 \times 10^{-2} eV  \gamma_{1,2}^4 n_1^{1/2}
\approx \cases { 2 \times 10^4 {\rm  MeV} f_4 \ef^2  n_1^{1/2}
&relativistic reverse shock  \cr\noalign{\smallskip}
2\times  10^8  {\rm  MeV}  \ef^4  n_1^{1/2}
&otherwise \cr}
\eqno(22)
$$
Similarly, the typical energy of a synchrotron photon emitted by
the reverse shock is:
$$
\epsilon_{o,rs} \approx 4 \times 10^{-2} eV  \gamma_{1,2}^2
\g_{3,4}^2 n_1^{1/2} \approx  2  {\rm  MeV}  \ef^2  n_1^{1/2}
\eqno(23)
$$
Interestingly enough this is in the right energy range regardless
of the question whether the reverse shock is relativistic or not.
This suggests that the observed radiation might come from the reverse
shock and demonstrates the potential of this mechanism. However it
also shows the difficulty that this mechanism poses.  Eqs. 22 and 23
depend on a relatively high power of $\eta$. For our canonical
parameters the radiation form the reverse shock is in the $\gamma$-ray
range. The load parameter, $\eta$ can, however, vary easily by orders
of magnitude from one burst to another. It is possible that similar
processes produce x-ray bursts, uv-bursts as well as bursts with much
harder $\g$-rays Alternatively, it is possible that the situation is
much more subtle and either $\eta$ is relatively constant or other
mechanisms control the emission. In either it is not clear yet
why does the energy emerge in soft $\g$-rays
and the resolution of this puzzle might provide the clue to the enigma
of grbs.

\heading {7. Beaming and  Timing}

The emitting source is moving relativistically towards  the observer
and the observed photons are blue shifted both in a
pure radiation fireball, that releases its photons
when it becomes optically thin, or in a loaded fireball that emits
a grb when it interacts with the ISM.
Thus,  each observer detects blue shifted
radiation from a narrow angle $\approx 1/\g$. This does not mean,
however, that the overall grb is beamed in such a narrow angle.  The
overall angular spread depends on the width of the
emitting region which depends on the source model
(in principle the emission could
be over $4 \pi$ if the fireball is spherically symmetric)
and is independent of $\g$.

The duration of the burst depends on several  factors. The original
duration of the pulse is $R_i/c \approx 10^{-3} \R07$.
This time scale increases due
to spreading of the pulse (which takes place in the free coasting
phase)  and could be as long as:
$$
\Delta T_1 \approx 1.5 \times 10^{-3}{\rm sec} \ \E051^{1/3} \ef^{-8/3}
n^{-1/3}
\eqno(24)
$$
for a loaded fireball with $R>R_w$ and $R>R_c$. The duration also
increases due to the small angular spread of the signal
A given observer will detect radiation from an angular scale
$1/\g$ around his line of sight. This will lead to a typical duration
of$^{21}$:
$$
\Delta T_2 \approx {R_\g / \gamma_F c}
= 5~ {\rm sec} \ \E051^{1/3} \ef^{-5/3} n^{-1/3}.
\eqno(25)
$$
The relatively strong dependence of $\eta$ is an advantage here, as
it provides a possible explanation to the large variability in durations
of grbs. Clearly $\Delta T$ is the longer of $\Delta T_1$ and $\Delta T_2$.

\heading{\bf 8. Conclusions}

We have shown  that fireballs with a large initial ratio $\e_i$ of
radiation energy to rest mass energy show certain common global
features during their expansion and evolution.  After a short initial
acceleration phase, the fluid reaches relativistic velocities, and the
energy and mass become concentrated in a radial pulse whose shape
remains frozen in the subsequent expansion.  The motion is then
described by an asymptotic solution (eqs 7, 10), which gives
for each individual shell scaling laws similar to those of a
homogeneous sphere.

The expanding fireball has two basic phases: a radiation dominated
phase and a matter dominated phase.  Initially, during the radiation
dominated phase the fluid accelerates with $\g \propto r$ for each
Lagrangian shell.  The fireball is roughly homogeneous in its local
rest frame but due to the Lorentz contraction its width in the
observer frame is $\Delta r \approx R_i$, the initial size of the
fireball.  Ultimately, a transition takes place to the matter
dominated phase and all the energy becomes concentrated in the kinetic
energy of the matter, and the matter coasts asymptotically with a
final Lorentz factor $\g_F \approx \eta$.  The matter dominated
phase is itself further divided into two sub-phases.  At first, there
is a frozen-coasting phase in which the fireball expands as a shell of
fixed radial width in its own local frame, with a width $\sim
\gamma_F R_i \sim \eta_i R_i$.  Because of Lorentz contraction the
pulse appears to an observer with a width $\Delta r \approx R_i$.
Eventually, the spread in $\g_F$ as a function of radius within
the fireball results in a spreading of the pulse and the fireball
enters the coasting-expanding phase. In this final phase, $\Delta r
\approx R_i / \g_F^2$, and the observed pulse width increases
linearly with the radius from which the radiation is emitted.

The fireball can become optically thin in any of the above phases.
Once this happen the system ceases to behave like a fluid, and the
radiation moves as a pulse with a constant width, while the baryons
enter a coasting phase like the one described above.

For most realistic loads the fireball becomes matter dominated before
it becomes optically thin (unless there is some unknown yet mechanism
that separates the baryons from the $e^+e^-$ pairs). In this case
the observed grb is produces at $R_\g \approx 10^{15}$cm from
the source where the kinetic energy of the baryons is converted to
thermal energy and $\gamma$-rays due to the interaction of the relativistic
baryons with the ISM. We have seen that this process leads to the right
time scales for grbs  and could potentially lead to the right spectrum.
However, at present the simple estimates of the spectrum do not
necessarily yield signals at the $\g$-ray range. An
explanations of this feature might provide the key to our understanding
of fireballs and grbs.

I would like to thank Ramesh Narayan for numerous helpful discussions.
This work was supported in part by a BRF grant to the Hebrew University and
NASA grant NAGS-1904 to the CFA.

\heading{Reference}
\def\ApJ{{\it Ap. J.}}
\def\ApJL{{\it Ap. J. L.}}
\def\Nature{{\it Nature}}
\def\etalc{{\it et. al., }}
\item
{1.} Meegan, C.A., \etalc \Nature, {\bf 355} 143.
\item
{2.} Meegan, C.A., \etalc this volume.
\item
{3.} Paczy\'nski, B. 1992, \Nature, 355, 521.
\item
{4.} Piran, T., 1992, \ApJL {\bf 389}, L45.
\item
{5.} Norris \etalc 1993, this volume.
\item
{6.} Davis \etalc 1993, this volume.
\item
{7.} Piran, T. and Shemi, A.,  1993, \ApJL, {\bf 403}, L67.
\item
{8.} Piran, T., 1993, This volume.
\item
{9.} Krolik,J.H. and Pier, E.A., 1991, \ApJ, {\bf 373}, 277.
\item
{10.} Leob, A., 1993, this volume.
\item
{11.} Dingus, B., \etalc 1993, this volume. 
\item
{12.} Goodman, J., 1986, \ApJL, {\bf 308} L47.
\item
{13.} Shemi, A. and Piran, T. 1990, \ApJL {\bf 365}, L55.
\item
{14.} Piran, T., Shemi, A. and Narayan, R., 1993,
{\it MNRAS.}, {\bf 263}, 861.
\item
{15.} Palmer, D., M., \etalc 1993, this volume. 
\item
{16.} Paczy\'nski, B., 1990. \ApJ, {\bf 363}, 218.
\item
{17.} Cavallo, G., and  Rees, M.J. 1978, {\it MNRAS.}, {\bf 183}, 359.
\item
{18.} Paczy\'nski, B., 1986, \ApJL, {\bf 308}, L51.
\item
{19.} M\'es\'zaros, P. \& Rees, M. J., 1992.
{\it MNRAS.}, {\bf 258}, 41p.
\item
{20.} M\'es\'zaros, P. \& Rees, M. J., 1992, \ApJL, in press.
\item
{21.} Katz, J. 1993, \ApJ in press.
\item
{22.} Blandford, R., D., and McKee, C. F., 1976, {\it Phys. of Fluids},
{\bf 19}, 1130.
\item
{23.} Blandford, R., D., and McKee, C. F., 1976,
{\it MNRAS.},  {\bf 180}, 343.
\item
{24.} M\'es\'zaros, P. \& Rees, M. J., 1993, this volume.
\bigskip
\heading{Figure Captions}
\item{Fig. 1} $R_\eta$ (solid line), $R_w$ (long dashed line),
$R_c$ (dotted line),
and $R_\g$ (short dashed line) as a function of $\eta$ for
$E_i=10^{51}$ergs and $R_i=10^7$cm.

\item{Fig. 2} Schematic density profile across the shocks. Region 1 is the
interstellar matter.  Regions 2 and 3 describe the shocked material,
with a contact discontinuity between 2 and 3.  Region 4 is the
unshocked material of the fireball.

\end